\documentclass[epj]{svjour}
\usepackage{graphics}
\begin{document}
\title{Ionization of atoms by few-cycle EUV laser pulses: carrier-envelope phase dependence of the intra-pulse interference effects}
\titlerunning{Carrier-envelope phase dependence of the intra-pulse interference effects.}
\subtitle{}
\author{Attila T\'{o}th\inst{1} \and S\'{a}ndor Borb\'{e}ly\inst{1} \and K\'{a}roly T\H{o}k\'{e}si\inst{2} \and Ladislau Nagy\inst{1} 
\authorrunning{Attila T\'{o}th \emph{et al.}}
}                     
%
\mail{sandor.borbely@phys.ubbcluj.ro}%
\institute{Faculty of Physics, Babe\c{s}-Bolyai University, Kog\u{a}lniceanu Street No. 1, 400084 Cluj-Napoca, Romania \and 
Institute for Nuclear Research, Hungarian Academy of Science, P.O. Box 51, H-4001 Debrecen, Hungary}
\date{Received: date / Revised version: date}
%
\abstract{
We have investigated the ionization of the H atom by intense few-cycle laser pulses, in particular the intra-pulse interference effects, 
and their dependence on the carrier-envelope phase (CEP) of the laser pulse. In the final momentum distribution of the continuum electrons the imprint of 
two types of intra-pulse interference effects can be observed, namely the temporal and spatial interference. During the spatial interference 
electronic wave packets emitted at the same time, but following different paths interfere leading to an interference pattern  measurable 
in the electron spectra. This can be also interpreted as the interference between a direct and a scattered wave, and the spatial interference 
pattern as the holographic mapping (HM) of the target. This HM pattern is strongly influenced by the carrier-envelope phase through the shape 
of the laser pulse. Here, we have studied how the shape of the HM pattern is modified by the CEP, and we have found an optimal CEP for the 
observation of HM.
\PACS{
      {PACS-key}{discribing text of that key}   \and
      {PACS-key}{discribing text of that key}
     } 
} 
\maketitle
\section{Introduction}
\label{intro}
With the progress of laser pulse generation techniques the production of few-cycle laser pulses for a wide range of photon energies 
became possible \cite{baltuska2003}. At high field intensities, when such an ultrashort laser pulse interacts with atoms, the dominant 
process is ionization. Beside ionization secondary processes also occur, which may have significant impact on the final distribution 
of the continuum electrons. Bian \emph{et al.} \cite{bian2011} recently showed that these secondary processes are the result of interference 
between electronic wave packets. From the numerous possible scenarios \cite{bian2011}, only two have a significant impact (measurable in 
experiments) on the final momentum distribution of the free electrons.

In the first scenario, electronic wave packets emitted at different time moments interfere, which leads to the formation of interference 
fringes in the electron energy spectrum consisting of concentric circular maxima and minima. This process is in fact a double-(multi-) 
slit interference in the time domain \cite{lindner2005}, and it was extensively studied both theoretically and experimentally.

In the second scenario, electronic wave packets emitted at the same time, but following different spatial paths interfere leading to a 
radial fringe structure in the electron energy spectrum \cite{bian2011,huismans2011,marchenko2011}. In a simplistic picture \cite{huismans2011} 
the radial fringe structure can be visualized as the interference between the direct and scattered electronic wave packets. By considering 
the direct wave packet as reference wave, while the scattered wave packet as a signal wave, the radial interference pattern can be interpreted 
as the holographic mapping (HM) of the target atom's state \cite{huismans2011}. In our previous work \cite{borbely2013,borbely2013a} we 
have showed the existence of two distinct types (direct and scattered) of electron trajectories which contribute to the formation of the 
HM interference pattern, and we have investigated in details how the laser field parameters are influencing the shape of the HM pattern. 
Based on the results of \cite{diego} in our previous study \cite{borbely2013} we have used a cosine-like laser pulse assuming that in this 
case the HM will be the dominant secondary process. In order to test this assumption, and to find the optimal conditions for the 
observation of the HM interference pattern, in the present work, we have performed several \emph{ab initio} calculations 
on the ionization of the H atom by differently shaped few-cycle laser pulses. The shape of a few-cycle laser pulse is controlled by the carrier-envelope phase (CEP), 
which measures the phase difference between the carrier (plane) wave and the envelope function. Our \emph{ab initio} calculations are based 
on the exact numerical solution of the time-dependent Schr\"odinger equation in the framework of the time-dependent close-coupling (TDCC) approach. \\
Throughout the present article, atomic units are used.

\section{Theory}
\label{sec:theory}
\begin{figure}
\resizebox{0.5\textwidth}{!}{%
  \includegraphics{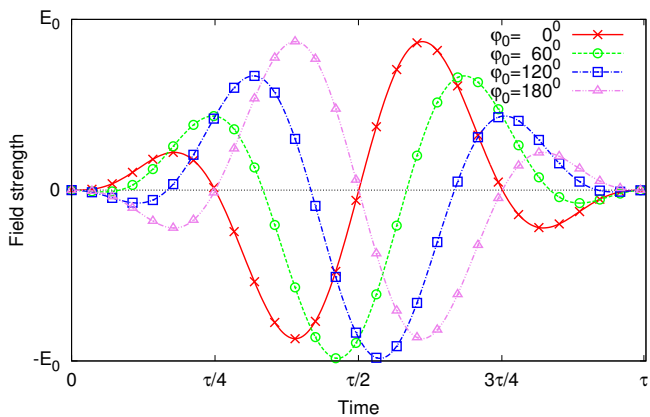}
}
\caption{The shape of the two-cycle laser pulses as a function of the $\varphi_0$ CEP.}
\label{fig:field}       
\end{figure}
The Hamiltonian describing our target (H atom) and its interaction with an external electromagnetic field can be written as
\begin{equation}
 \hat H = \frac{\hat p^2}{2}+V(\vec{r})+\vec{r}\cdot\vec{E}(t).
\end{equation}
Here $\hat p^2/2$ is the electron's kinetic energy operator, $V(\vec{r})=-1/r$ is the Coulomb interaction potential between the core and electron, while 
$\vec{r}\cdot\vec{E}(t)$ is the interaction between the electron and the external laser field expressed in length gauge and 
in dipole approximation. The external laser pulse is characterized by its electric component, which is a plane wave 
modulated by a sine-square envelope function:
\begin{equation}
 \vec{E}(t)=\left\{ 
 \begin{array}{ll}
  \hat\varepsilon E_0\sin^2\left(\frac{\pi t}{\tau}\right)\sin(\omega t -\varphi_0) & \mathrm{if}\ t \in (0,\tau)\\
  0 & \mathrm{elsewhere}
 \end{array}
 \right.,
\end{equation}
where $\hat\varepsilon$ is the polarization of the laser field pointing in the Oz direction, $\omega$ is the frequency of the carrier wave, $\varphi_0$ is the 
carrier-envelope phase, and $\tau$ is the pulse duration. As it is illustrated on Fig \ref{fig:field}, the CEP significantly 
influences the shape of a the electric field of a two-cycle laser pulse, and through this the electron dynamics driven by the 
laser field.

\subsection{The TDCC model}
\label{sec:tdcc}
The time evolution of a hydrogen atom driven by an external electric field is described by the time-dependent Schr\"odinger equation
\begin{equation}
 i\frac{\partial}{\partial t} \Psi(\vec{r},t)=\hat H\Psi(\vec{r},t),
 \label{eq:tdse}
\end{equation}
where $\Psi(\vec{r},t)$ is the electron's time-dependent wave function. In the present work for the solution of Eq (\ref{eq:tdse}) 
we have used the time-dependent close-coupling (TDCC) method \cite{colgan}, where the time-dependent wave function is expanded in the 
basis of spherical harmonics
\begin{equation}
 \Psi(\vec{r},t)=\sum\limits_{lm}\frac{R_{lm}(r,t)}{r}Y_{lm}(\Omega_r).
 \label{eq:exp}
\end{equation}
By substituting the expansion (\ref{eq:exp}) into Eq. (\ref{eq:tdse}) we obtain a set of coupled differential equations, the 
time-dependent close-coupling equations, for the $R_{lm}(r,t)$ radial wave functions (for details see \cite{borbely2013}). The TDCC 
equations are solved (i.e. the $R_{lm}(r,t)$ wave functions are propagated in time) using the short-iterative Arnoldi-Lanczos method 
\cite{park}, which is a unitary, and unconditionally stable time propagation method. For the discretization of the radial wave functions 
we have used the finite-element discrete variable representation (FEDVR) method \cite{schneider}, where the radial coordinates are 
divided into finite elements (FE), and inside each FE the wave function is represented on a local discrete variable representation 
(DVR) basis. To ensure the continuity of the wave functions at the FE boundaries, the local DVR basis was built on top of Gauss-Lobatto 
quadrature points. After the end of the laser pulse the distribution of the continuum electrons (i.e ionization spectrum) is calculated 
by the direct projection of the time-dependent wave function (\ref{eq:exp}) into exact continuum eigenstates (i.e. into Coulomb wavefunctions) 
of the H atom.

\section{Results}
\label{sec:results}
In order to investigate the CEP dependence of the HM interference pattern we have performed calculations for the ionization of the H atom by 
two-cycle laser pulses at two field intensities $E_0\in \{0.37~\textrm{a.u.}, 1.00~\textrm{a.u.}\}$ for several CEP values 
$\varphi_0\in \{0^\circ, 30^\circ, 60^\circ, 90^\circ, 120^\circ, 150^\circ, 180^\circ \}$.
\begin{figure*}
\centering
\resizebox{0.95\textwidth}{!}{%
  \includegraphics{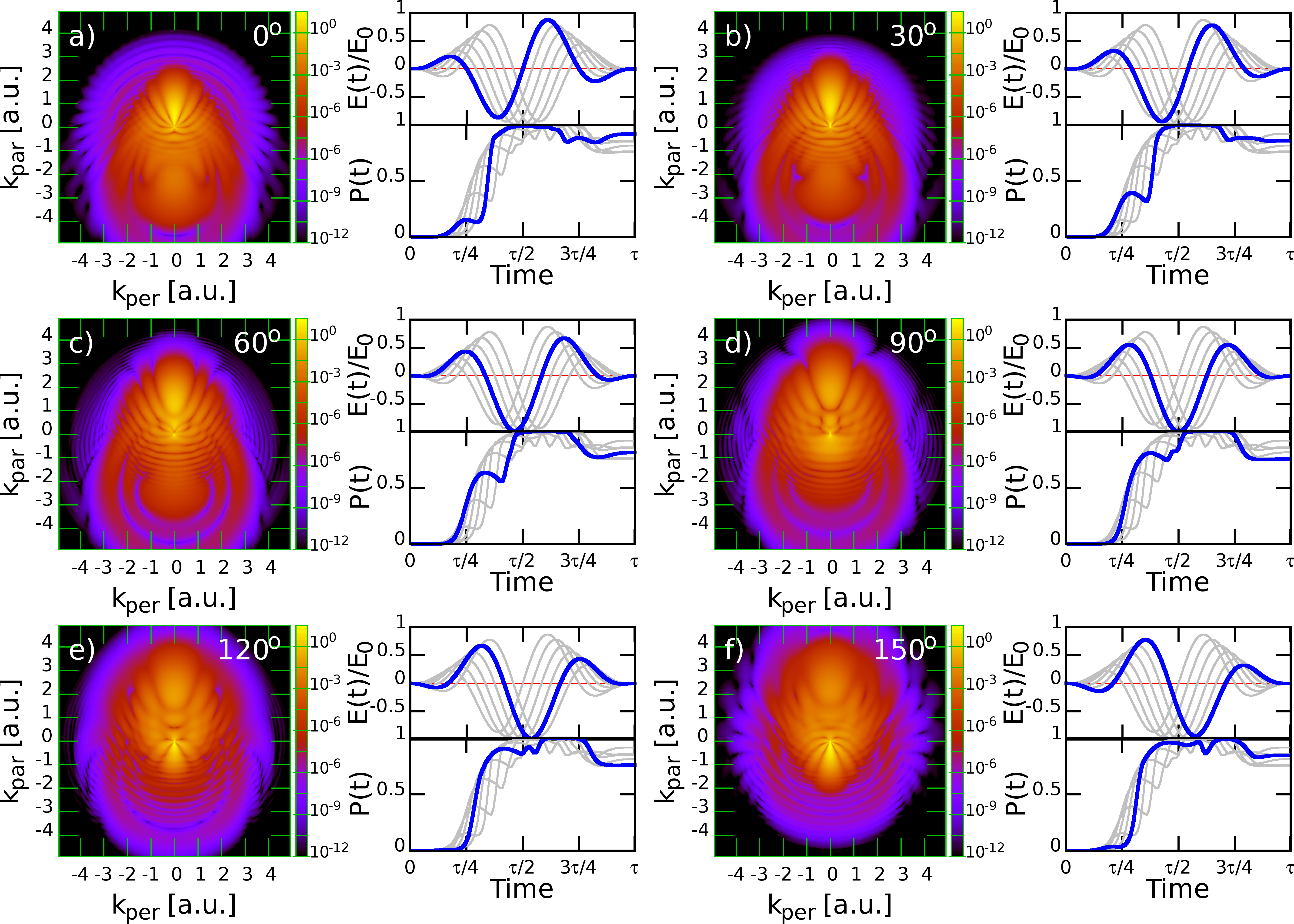}
}
\caption{The momentum distribution of the continuum electrons as function of perpendicular ($k_{per}$) and parallel ($k_{par}$) momentum 
components for $E_0=1.00~$ a.u. field intensity, and for the following CEP values: a) $\varphi_0=0^\circ$; b) $\varphi_0=30^\circ$, 
c) $\varphi_0=60^\circ$, d) $\varphi_0=90^\circ$, e) $\varphi_0=120^\circ$, f) $\varphi_0=150^\circ$. On the right hand side of each momentum 
distribution the corresponding field strength $E(t)$, and total ionization probability $P(t)$ is plotted with thick lines as a function 
time.}
\label{fig:e100}       
\end{figure*}
During the calculations the frequency of the carrier wave ($\omega_0 = 0.4445~$ a.u.) and the pulse duration ($\tau = 28.27~$a.u. corresponding to 
two field oscillations) was fixed. The obtained momentum distributions of the continuum electrons are plotted on Fig. \ref{fig:e100} for 
$E_0 = 1.00~$a.u., and on Fig. \ref{fig:e037} for $E_0 = 0.37~$a.u.. On the right hand side of each momentum map the corresponding strength of the 
laser pulse $E(t)/E_0$, and the total ionization probability $P(t)$ as a function of time are plotted with a thick line. 
In order to make the comparison easier, these plots also present the $E(t)/E_0$ and $P(t)$ curves (thin lines) obtained for the other CEP values.
Since the $\varphi_0 = 180^\circ$ results can be obtained from the $\varphi_0 = 0^\circ$ ones by performing a simple $k_{par}\rightarrow -k_{par}$ 
mirror transformation, on Figs. \ref{fig:e100}-\ref{fig:e037} we omitted them.

At both field strength, regardless of the value of the CEP, one can clearly identify the HM interference patterns as the quasi radial minima 
and maxima ridges in the momentum distribution plots. 
Another similarity of all sets of results is that ionization occurs in consecutive bursts around the peaks of the oscillating electric field indicating that the dominant mechanisms are the tunneling and over-the-barrier ionization. 
The position and instantaneous strength of these peaks is greatly influenced by the CEP, 
and ultimately affects both the shape and the visibility of the resulting HM interference pattern. In case of laser pulses containing 
many optical cycles of the carrier wave, there are a number of peaks with similar instantaneous intensities and the effect of the CEP is smeared out.

\begin{figure*}
\centering
\resizebox{0.95\textwidth}{!}{%
  \includegraphics{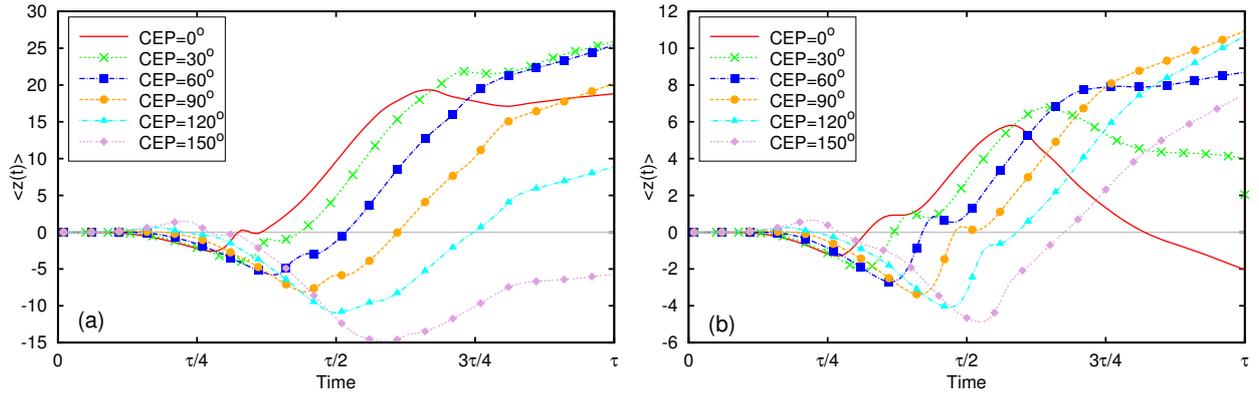}
}
\caption{Expectation value of the $z$ coordinate for different CEP values of the two-cycle laser pulse: a) $E_0=1.00$~a.u.; b)$E_0=0.37$~a.u..}
\label{fig:zt}       
\end{figure*}

Beside the momentum maps we also calculated the expectation value of $z$ coordinate as a function of time (see Fig \ref{fig:zt}), which gave us further insight into the wave packet dynamics, and consequently better understand the role of the CEP for a clear HM interference pattern.
\begin{figure*}
\centering
\resizebox{0.95\textwidth}{!}{%
  \includegraphics{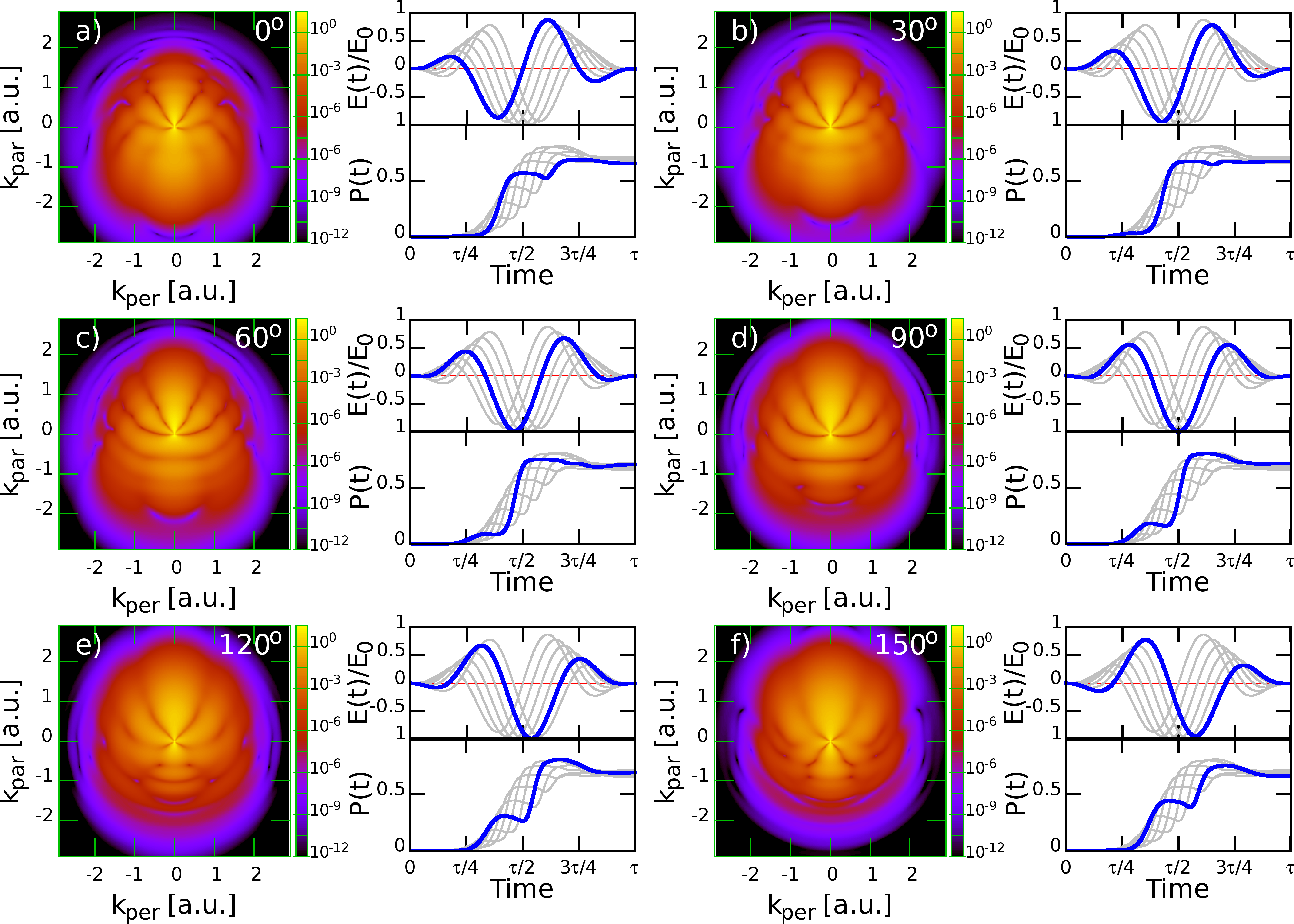}
}
\caption{Same as Fig \ref{fig:e100} but for $E_0=0.37~$a.u..}
\label{fig:e037}       
\end{figure*}

By taking a closer look at Fig. \ref{fig:e100} we can see that the HM structures are the clearest for the $\varphi_0=60^{\circ}$ and $\varphi_0=90^{\circ}$ 
(cosine-shaped pulse) calculations. Also, the temporal interference rings \cite{borbely2013} are best identifiable for these CEP values. This seems to be in contradiction 
with the findings of \cite{diego}, namely that the temporal interference should be the dominant feature of the momentum distributions obtained after the interaction 
with a sine-shaped ($\varphi_0=0^{\circ}$) laser pulse. The absence of dominant temporal interference fringes from our momentum maps can be attributed to the laser pulse parameters used in our calculations. For the formation of prominent temporal interference fringes it is an essential requirement that at least two continuum electronic wave packets with comparable stengths to be emitted at different time moments. In Figs. \ref{fig:e100}-\ref{fig:e037} for our sine-shaped pulses we can observe the formation of only one dominant continuum wave packet (only one sudden increase in the P(t) total ionization probability), thus the conditions for temporal interference are not fulfilled.

In the case of the clearest HM pattern ($\varphi_0=90^{\circ}$), a free wave packet $\psi_{c1}$ ($\approx80\%$ of the entire population) 
is created by the first peak of the laser pulse near $\tau/4$. This is first accelerated in the $-z$ direction, than it is turned around by the second peak of 
the pulse centered around $\tau/2$, and scatters on the residual ion. During this second peak a small portion of the returning wave packet is recaptured followed by the emission of another wave packet $\psi_{c2}$, leaving the target nearly depleted. At this stage both wave packets propagate in the positive $z$ direction. When the laser field changes its sign again (the peak near $3\tau/4$) the free wave packets are slowed down, and their low momentum components which did not depart too far are brought back to the near vicinity of the core, where based on the evolution of the total ionization probability (the sudden drop in P(t) around $3\tau/4$) they are recaptured. By considering that the $\psi_{c2}$ wave packet was ``born'' in the raising edge of the second field peak, and that the strength of the third field peak (located near $3\tau/4$) is much lower than the strength of the second one, we can conclude that the $\psi_{c2}$ wave packet will not recollide with the core. Thus in the momentum map the HM interference pattern is the result of a single recollision of the $\psi_{c1}$ wave packet. A similar wave packet dynamics is observed for the $\varphi_0=60^\circ$.  This dynamics is also illustrated on Fig \ref{fig:zt}, where the expectation value of the $z$ 
coordinate is plotted as a function of time. Due to the fact that the quiver motion of the free electrons is along the laser polarization axis ($z$ direction),
these figures provide good insight into the continuum electron dynamics, but they are not able to resolve the separate motion of the different ($\psi_{c1}$ and $\psi_{c2}$) wave packets.

In contrast to the above discussed cases ($\varphi_0=60^\circ$ and $\ 90^\circ$), for all the other CEP values during the dynamics multiple rescattering events are present. As a result of this in the momentum maps several (at least two) HM patterns are present simultaneously, which makes the study of their properties cumbersome. This is illustrated by the $\varphi_0=30^\circ$ case, where besides the $\psi_{c1}$ wave packet the second wave packet also recollides with the core. This is possible because for $\varphi_0=30^\circ$ the strength of the second and third field peaks is roughly the same. A different scenario can be observed for $\varphi_0=120^\circ$, where we have one dominant continuum wave packet emitted at the very beginning of the laser field (at $\tau/4$), which then recollides with the core twice. This multiple recollisions are also shown on Fig. \ref{fig:zt}, where the $\varphi_0=120^\circ$ $z(t)$ curve crosses the $z=0$ line twice. In all scenarios as a result, both the positive 
and the negative $k_{par}$ half of the $\{ k_{par}, k_{per}\}$ configuration space contains spatial interference structures. 

By changing the CEP, the shape of the laser pulse, i.e. the relative intensities of consecutive peaks, is also modified. This strongly influences the $z_0$ distances reached by the free wave packets between consecutive redirections. The shape (i.e density) of the HM pattern is directly influenced by the $z_0$ distance reached by the free wave packets before recollision (for details see \cite{borbely2013,borbely2013a}). This is the reason why the density of the HM pattern depends so strongly on the CEP value.

Our calculations for $E_0=0.37~$a.u. show a similar behavior, although the exact evolution of the system is somewhat different due to the lower 
intensities considered. During these calculations, the target is not fully depleted, and as a result the recapture of the continuum electrons at 
the end of the pulse is not as pronounced as in the higher intensity case. Also, as we have seen in our previous studies (\cite{borbely2013,borbely2013a}) lower intensity means that the free wave packets gain smaller average velocities, thus reach smaller distances from the 
scattering center between consecutive redirections, which manifests in less detailed (e.g. fewer interference extrema) HM patterns. Nevertheless, 
the $\varphi_0=60^{\circ}$ and $\varphi_0=90^{\circ}$ results still present the cleanest interference patterns again. For the other considered CEP 
values, just as in the $E_0=1.00~$a.u. case, the momentum distribution is smeared by the overlap of two HM structures created by consecutive 
scattering events. Another difference compared to the higher intensity case is that for these calculations there are no pronounced temporal interference 
rings present in the momentum distributions.

\section{Conclusions}
\label{sec:conclusions}

In the present work we have investigated theoretically the ionization of atomic hydrogen by intense two-cycle 
laser pulses. Our principal research interests lie in the description of the intra-pulse interference effects, 
most notably the spatial interferences which can be interpreted as the holographic mapping of the target. Here, 
we investigated how the carrier envelope phase influences these interference patterns in the momentum distribution 
of the continuum electrons. We have seen that for the considered few-cycle radiations the CEP has a huge impact 
on the shape of the laser pulse, and implicitly on the dynamics of the system. Based on our results we can 
conclude that in order to obtain a clear HM interference pattern it is desirable that only one dominant scattering 
to take place. This condition is fulfilled for laser pulses with large asymmetry of the neighboring pulse peak 
strengths, which is the case predominantly in the $60^{\circ}\leq \varphi_0 \leq 90^{\circ}$ CEP region.

\begin{acknowledgement}
 This work was supported by a grant of the 
Romanian National Authority for Scientific Research, CNCS UEFISCDI Project No. PN-II-ID-PCE-2011-3-0192, 
by the European Cost Action CM1204 (XLIC), and partially the Hungarian Scientific Research Fund OTKA Nos. NN103279 and K103917.
\end{acknowledgement}

%
%

\end{document}